\documentstyle[preprint,aps]{revtex}
\begin{document} \draft
\title{Magnetotunneling
through a semiconductor double barrier structure assisted by light.}
\author{Jesus I\~{n}arrea and  Gloria Platero}
\address{Instituto de Ciencia de Materiales (CSIC) and  Departamento de
Fisica de la Materia Condensada C-III, Universidad Autonoma,
Cantoblanco, 28049 Madrid, Spain.}
\maketitle
\begin{abstract}
We have analyzed theoretically the effect of a laser on the tunneling
current through a double barrier in the presence of a parallel magnetic
field. The magnetotunneling current is modified by the light due to
the photon emission and absorption processes which assist the tunneling
of electrons through the structure. We observe that the effect of the
light can be controlled by tuning the ratio between the cyclotron and
the photon frequencies, turning out that when this ratio is one  the
effect of the light is drastically reduced. The change of the
accumulated charge into the well due to the light and therefore the
modification of the number of Landau levels contributing to the
current is discussed.
\end{abstract}
\newpage
\narrowtext
\section{INTRODUCTION}
\par
The analysis of resonant tunneling through semiconductor
heterostructures has a lot of interest from a fundamental point of view
as well as for  its applications as microelectronic devices \cite{1}.
 From the analysis of the tunneling current as a function of an external
DC bias many properties of the electronic band structure
and of the excitations in the
system can be deduced. The application of an external field increases the
parameters which can be externally controlled and gives additional
information of the heterostructure and
therefore increases the applications of
these systems as devices.
 The analysis of the magnetotunneling current  for both configurations,
parallel and perpendicular to the growth direction of the
heterostructure has been the subject of
 many papers \cite{2,3,4}
  and  gives
information on the density of states in the well corresponding to the
Landau level ladder, in the first case and on the edge states in the
second configuration.
 Also the application of a radiation field modifies the transport
properties of these resonant devices.\\
The work of Sollner et al\cite{5}, is the experimental starting
point for
studies on the effect of time-dependent potentials in resonant tunneling
through semiconductor microstructures: they studied the influence of
electromagnetic
radiation on resonant tunneling current. Recently Chitta et al \cite{6}
have studied the far infrared response of double barrier
resonant tunneling
structures.
Tien and
Gordon \cite{7}, studied the effect that microwave radiation has on
superconducting tunneling devices.
 Several authors \cite{8,9,10,11,12}
have investigated the effect that external  time dependent potentials have in
different problems.
 In a recent work \cite{13,14} we have developed a model to
analyze the coherent and sequential tunneling current in the presence
of a photon field through a double barrier structure.
\par
 In this
paper we have analyzed the effect of a photon field on both coherent
and sequential magnetotunneling current through a DBS for a static
magnetic  field
applied in the direction of the current.
We have developed a quantum mechanical formalism to find the expression
for the electronic state dressed by photons and
 we propose a  model to obtain
the sequential  magnetotunneling current under the influence of an external
electromagnetic field for linearly  polarized light. By means
of this model we obtain the charge into the well and its modifications
due to the external electromagnetic field.\\
\par
In our model for sequential magnetotunneling , before switching on the
light, we consider that
electrons tunnel coherently through the first barrier, then they loss
memory within the well and in a third step they cross coherently the
second barrier. In the presence of a magnetic field,  coherent tunneling
implies Landau level index conservation, however , for sequential
tunneling, due to the scattering
processes which take place into the well structure, the Landau level
index corresponding to the states which contribute to the current is not
conserved from the emitter to the collector.
 If a laser is switched on, it
changes the charge occupation in the well
 and consequently the number of
 Landau levels (LL)  which contribute to the current density .
We have analyzed this effect for different
 photon frequencies and field
intensities as well as the change of the magnetotunneling current for
different
ratios between  cyclotron and  photon frequencies.
 We have compared also
the two contributions to the magnetotunneling current:
 coherent versus sequential. \\
\par
This paper is organized as follows: In Sec II, we discuss and
develop the theoretical
 formalism for the coherent and sequential magnetotunneling in
the presence of a photon field .
In Sec III, we applied the discussed formalism to analyze both
contributions to the current: sequential and coherent for different
ratios between the cyclotron and the photon frequencies.
Also the accumulated charge into the well for
 different magnetic and electromagnetic fields is analyzed for
sequential magnetotunneling and therefore how the light changes the
number of Landau levels which contribute to the sequential tunneling.
We summarize our conclusions in Sec IV.\\

\section{ MAGNETOTUNNELING CURRENT TROUGH A DOUBLE BARRIER STRUCTURE
IN THE PRESENCE OF AN ELECTROMAGNETIC FIELD.}
\subsection{Coherent magnetotunneling}
\par
 We have analyzed the coherent magnetotunneling current
 density through a double barrier when a magnetic field
is applied parallel to the current in the presence of light .
 We have considered light linearly
polarized in the far infrared regime, and polarized in the same
direction as the static magnetic field applied . For that configuration, the
electronic motion is modified
 by the light only in the current direction
and the electronic lateral state becomes unaffected by light \cite{13,14}.
With no magnetic field present in the sample the parallel component for
the electronic wave vector is conserved during the coherent
tunneling process. Now as the magnetic field is switched on,
is the Landau level index
what is conserved:
the current characteristic curve presents a peak as a
function of the external bias when a Landau level in the emitter
aligns with the corresponding Landau level in the well. As the magnetic
field increases, less Landau levels contribute to the current, but the
degeneration of each LL increases, giving in the current less peaks
but more intense.\\
The effect of a magnetic field $\vec{B}$, parallel to the current direction,
i.e., the $z$
direction, is to change the
parallel part of the density of states and due to that instead
of a continuum of states we have now a Landau levels
ladder. If  a laser is applied to the sample, the Hamiltonian for an
electron in the presence of an
electromagnetic field in the configuration considered above and
a magnetic field parallel to the current can be describe in
second quantization as:
\begin{eqnarray}
H_{tot}=H_{e}^{0}+H_{ph}^{0}+W_{D}(t)+W_{OD}(t)
\end{eqnarray}
where
\begin{eqnarray}
H_{e}^{0}&=&\sum_{k} \epsilon_{z} c_{z}^{+} c_{z}+
     \hbar w_{c}(a_{B}^{+} a_{B} + 1/2) \\
H_{ph}^{0}&=&\hbar w a^{+}a\\
W_{D}(t)&=&\sum_{k} [(e/m^{*})<k|P_{z}|k> c_{k}^{+} c_{k}
  (\hbar/2\epsilon Vw)^{1/2} (a e^{-iwt} +a^{+} e^{iwt})]\\
W_{OD}(t)&=& \sum_{k}\sum_{k^{'}\neq k} [(e/m^{*})<k^{'}|P_{z}|k>
  c_{k^{'}}^{+} c_{k} (\hbar/2\epsilon Vw)^{1/2}
   (a e^{-iwt} + a^{+} e^{iwt})]
\end{eqnarray}
where $B$  is the magnetic field intensity,
$w_{c}$  is the cyclotron frequency: $w_{c}=eB/m^{*}$,
$a_{B}^{+}$ and $a_{B}$ are the creation and destruction
operators for the Landau states, and $\epsilon_{z}$ is
the electronic energy perpendicular part and
$A_{z}(x,t)=(\hbar/2\epsilon Vw)^{1/2}
\vec{\varepsilon_{z}} (a e^{-iwt} + a^{+} e^{iwt}) $ being $w$ the photon
frequency ( the wave vector of the electromagnetic
field has been neglected) . $H_{e}^{0}$ is the independent
electronic Hamiltonian and includes the double barrier
potential and the external applied bias, therefore
the eigenstates of $H_{e}^{0}$, $\Psi_{0}(k)$, are the tunneling states
for bare electrons in the presence of a magnetic field.
$H_{ph}^{0}$ is the photon field Hamiltonian without coupling with
electrons and $W_{D}$ and $W_{OD}$, describe the coupling
between electrons
and photons in the total Hamiltonian.
 Following \cite{13,14} we separate the coupling term in the
"diagonal"  and the "off-diagonal" contributions:
\begin{eqnarray}
H_{tot}=H_{D}(t)+W_{OD}(t)
\end{eqnarray}
where $H_{D}(t)=H_{e}^{0}+H_{ph}^{0}+W_{D}(t) $
\\
\par
The hamiltonian $H_{D}$, can be solved exactly
considering a canonical transformation\cite{15} and the off-diagonal
term is treated in time dependent perturbation theory using the same
procedure as in refs. [13,14] for coherent resonant tunneling .
The expression for the coherent magnetotunneling
current can be written then as:
\begin{eqnarray}
J=(2/2\pi^{2})(e/\hbar)^{2} B \sum_{n=0}^{N}\int_{(n+1/2)\hbar w}
 ^{E_{F}} dE [f(E)-f(E+V_{f})] T(E,n)
\end{eqnarray}
being $n$ the Landau level index, $N$, the maximum occupied Landau
level index, and $T(E,n)$ the  coherent transmission
coefficient through a double barrier structure
 when the photon field is present in the sample\cite{13,14}.\\
\par
\subsection{Sequential magnetotunneling}
\par
In order to describe the effect of the light on the tunneling current
and to compare with experiments,
one should also analyze how the sequential contribution to the tunneling
current is affected by the light.\\
 The electrons lose coherence when tunnel through the
structure due to the different scattering processes which suffer
with impurities, surface roughness, phonons..
Once the electrons cross the first barrier and if the scattering
time is shorter than the tunneling time they relax into the well
losing memory and in a next step they cross coherently the
collector barrier.\\
In order to study the sequential tunneling current ,
before illuminating the sample, we have calculated in
the framework of the Transfer Hamiltonian formalism
 the current through the first and the
second barriers separately, $J_{1}$, and $J_{2}$. These currents are
related to the Fermi level in the well $E_{w}$ or in
other words, to the amount of electronic charge stored into
the well. In this model we adjust selfconsistently the Fermi
level into the well
invoking current conservation through the whole
heterostructure.
The values calculated in this way for
the current and the Fermi level in the well, are indeed the actual current
which is crossing the whole double barrier sequentially and the Fermi level
corresponding to the actual amount of charge  stored
into the well.
We improve a previous model for sequential
tunneling\cite{13} considering instead of a discrete
level into the well ,a
localized state with finite width due to its coupling with the
continuum of states in the leads.
 This model takes into account macroscopically the
possible
scattering processes within the well.\\
The expression for the current through the emitter barrier $J_{1}$ to the
resonant state in the well can be written including the finite width of
the resonant well state as:
\begin{eqnarray}
J_{1}=\frac{e}{2 \pi^{2} \hbar} \int_{0}^{E_{F}} \frac{k_{w} T_{1}
L(E_{z} - E_{tn})}{w_{2}+1/\alpha_{e}+1/\alpha_{c}} (E_{F}-E_{z}-E_{w})
 dE_{z}
\end{eqnarray}
Where $k_{w}$ is the electronic perpendicular wave vector in the well,
$T_{1}$ is the single barrier transmission coefficient for the first
barrier, $w_{2}$ is the well width, $\alpha_{e}$ and $\alpha_{c}$ are
the perpendicular electronic wave vectors in the emitter and collector
barriers respectively, $L(E_{z}-E_{tn})=\frac{\gamma}{\pi
[(E_{z}-E_{tn})^{2}+ \gamma^{2}]}$, $\gamma$ is the half width
of the resonant state, $E_{tn}$ is the resonant well state
energy referred to the conduction band bottom  and
$E_{w}$ is the chemical potential in the well.
The current through the collector barrier $J_{2}$  for electrons coming
from the well  can be evaluated in the same way:
\begin{eqnarray}
J_{2}=\frac{e}{2 \pi^{2} \hbar} \int_{0}^{E_{F}} \frac{k_{w} T_{2}
L(E_{z} - E_{tn})}{w_{2}+1/\alpha_{e}+1/\alpha_{c}} E_{w}
 dE_{z}
\end{eqnarray}
where $T_{2}$ is the single barrier transmission coefficient for the
the second barrier (collector barrier).
Applying the initial condiction of $J_{1}=J_{2}$, we can obtain analytically
an expression for the total current which is crossing sequentially the
DB without light present:
\begin{eqnarray}
J_{T}=\frac{e}{2 \pi^{2} \hbar} \int_{0}^{E_{F}} \frac{k_{w}
L(E_{z} - E_{tn})}{w_{2}+1/\alpha_{e}+1/\alpha_{c}} (E_{F}-E_{z})
\frac{T_{1} T_{2}}{T_{1}+T_{2}} dE_{z}
\end{eqnarray}
In the presence of light the sequential current can be evaluated
within the framework of time dependent perturbation theory as in
the case of coherent tunneling for each barrier\cite{13,14},
including the finite width of the
resonant state. The expression obtained
for the total sequential current through the DB invoking current
conservation through the structure is:
\begin{eqnarray}
J_{T}&=&\frac{e}{2 \pi^{2} \hbar} \int_{0}^{E_{F}}
\frac{k_{w}(E_{F}-E_{z}) T_{2t}}
{w_{2}+1/\alpha_{e}+1/\alpha_{c}}\nonumber\\
     & &[L(E_{z}-E_{tn}) \frac{T_{d} }{ T_{d}+T_{2t}} +
L(E_{z}+ \hbar w -E_{tn}) \frac{T_{a} }{T_{a}+T_{2t}} + \nonumber\\
     & &L(E_{z}-\hbar w - E_{tn}) \frac{T_{e}}{T_{e}+T_{2t}}] dE_{z}
\end{eqnarray}
where:
\begin{eqnarray}
T_{d}&=&\frac{T_{1}(E_{z})} {1+k_{1}/k_{0}|C_{1,0}^{(1)}|^{2}+
k_{-1}/k_{0}|C_{-1,0}^{(1)}|^{2}}\\
T_{a}&=&\frac{T_{1}(E_{z}+\hbar w) |C_{1,0}^{(1)}|^{2}}
{k_{0}/k_{1}+|C_{1,-0}^{(1)}|^{2}+k_{-1}/k_{1}|C_{-1,0}^{(1)}|^{2}}\\
T_{e}&=&\frac{T_{1}(E_{z}-\hbar w) |C_{-1,0}^{(1)}|^{2}}
{k_{0}/k_{-1}+|C_{1,0}^{(1)}|^{2}k_{1}/k_{-1}+|C_{-1,0}^{(1)}|^{2}}\\
\end{eqnarray}
and $T_{1t}=T_{d}+T_{a}+T_{e}$ and $T_{2t}$ are the transmission coefficients
for the single barriers (emitter and collector respectively) in the
presence of the photon field:
\begin{eqnarray}
T_{2t}&=& \frac{T_{2}(E_{z})} {1+k_{1}/k_{0}|C_{1,0}^{(1)}|^{2}+
k_{-1}/k_{0}|C_{-1,0}^{(1)}|^{2}}+\nonumber\\
      & &\frac{T_{2}(E_{z}+\hbar w) |C_{1,0}^{(1)}|^{2}}
{k_{0}/k_{1}+|C_{1,-0}^{(1)}|^{2}+k_{-1}/k_{1}|C_{-1,0}^{(1)}|^{2}}+
 \nonumber\\
      & &\frac{T_{2}(E_{z}-\hbar w) |C_{-1,0}^{(1)}|^{2}}
{k_{0}/k_{-1}+|C_{1,0}^{(1)}|^{2}k_{1}/k_{-1}+|C_{-1,0}^{(1)}|^{2}}
\end{eqnarray}
$C_{i,j}$ are the coefficients of the wave functions :
\begin{eqnarray}
\Psi_{0}(t)&=&\alpha[\Phi_D(k_{0}) +
C_{1,0}\Phi_D(k_{1}) e^{-iwt}+\nonumber\\
           & &C_{-1,0}\Phi_D(k_{- 1})e^{iwt}] e^{-iw_{0}t}\\
\Psi_{1}(t)&=&\alpha^{'}[\Phi_D(k_{1})e^{-iwt} +
C_{1,1}\Phi_D(k_{2})e^{-2iwt}+\nonumber\\
           & &C_{-1,1}\Phi_D(k _{0})] e^{-iw_{0}t}\\
\Psi_{-1}(t)&=&\alpha^{"}[\Phi_D(k_{-1})e^{iwt}
 + C_{1,-1}\Phi_D(k_{0}) +\nonumber\\
            & &C_{-1,-1}\Phi_D(k_{- 2})e^{2iwt}]e^{-iw_{0}t}
\end{eqnarray}
 Here, $\alpha$, $\alpha^{'}$ and $\alpha^{"}$ are normalization
constants ,$\Psi_{0}$
corresponds to a state at one photon
 energy lower ( higher) than $\Psi_{1}$
($\Psi_{-1}$), and the coefficients $C_{i,j}$ are
 given in ref [13]. For
those coefficients, the first subscript is referred  to the interaction
with light processes i.e., '1'('-1') means absorption (emission) process,
whereas the second subscript is referred to the state i.e.,
 the subscript "0" means the reference  state energy
and the subscripts  "1" and "-1"
mean one photon energy above and below that reference
 state respectively.\\
\par
We will consider now the sequential magnetotunneling current: before
switching on the light the electrons tunnel sequentially through the
first and second barrier suffering scattering events into
the well as it was previously discussed.
 In this case, the LL index conservation takes
place from the first barrier to the well and from there to the collector
independently and not through the whole structure as
in the case of coherent tunneling.\\
The expression for the current through the first barrier $J_{1}$ before
illuminating the sample and in the presence of a
magnetic field can be written as :
\begin{eqnarray}
J_{1}=\frac{e^{2} B}{2\pi^{2} m^{*}}\int_{0}^{E_{F}-1/2\hbar w_{c}}
\frac{k_{w} T_{1}L(E_{z}-E_{tn})}{(w_{2}+1/\alpha_{e}+1/\alpha_{c})}
(N_{t}-\Delta) dE_{z}
\end{eqnarray}
 where $\Delta$ is the fraction of the total Landau levels which is
occupied in the well, $N_{t}$ the total number of Landau levels available
to tunnel for an external applied bias
( $\Delta$ runs between 0 and $N_{t}$),
$w_{c}$ is the cyclotron frequency and $w_{2}$ the well thickness.
 For the second barrier, we apply exactly the same
formalism and we obtain for the current through the second barrier $J_{2}$:
\begin{eqnarray}
J_{2}=\frac{e^{2} B}{2\pi^{2} m^{*}}\int_{0}^{E_{F}-1/2\hbar w_{c}}
\frac{k_{w} T_{2}L(E_{z}-E_{tn})}{(w_{2}+1/\alpha_{e}+1/\alpha_{c})}
\Delta dE_{z}
\end{eqnarray}
The sequential magnetotunneling
current is obtained when both currents $J_{1}$ and
$J_{2}$ are equal and it determines the Landau levels
which are  occupied within the
well and therefore the Fermi energy into the well. Doing this way
we can obtain an analytical expression for the total
magnetocurrent which
crosses sequentially the DB before illuminating the
sample:
\begin{eqnarray}
J_{T}=\frac{e^{2} B}{2\pi^{2} m^{*}}\int_{0}^{E_{F}-1/2\hbar w_{c}}
\frac{k_{w} L(E_{z}-E_{tn})}{(w_{2}+1/\alpha_{e}+1/\alpha_{c})}
\frac{T_{1} T_{2}}{T_{1}+ T_{2}} \Delta dE_{z}
\end{eqnarray}
Before switching on the light the electrons in the emitter  have just
one way to tunnel resonantly
into the well : from an emitter state which is resonant with the well state.
Once the external electromagnetic field is applied to the structure,
there is a coupling between the photons and the electrons tunneling
through the barrier structure.
Now the electrons have three different ways
to tunnel through the emitter barrier to the well. The
first one is a direct way and corresponds to an emitter state which
resonates with the well state,i.e., the transmission takes place without light
absorption or emission. The second one is through and absorption process
from an emitter state which is found at one photon energy below the resonant
well state and finally the third way is through an emission process from
an emitter state which is found at one photon energy above the resonant
well state. After some algebra , we obtain the next expression
for the sequential magnetotunneling current assisted by light:
\begin{eqnarray}
J_{T}&=&\frac{e^{2} B}{2 \pi^{2} m^{*}} \int_{0}^{E_{F}-1/2 \hbar w_{c}}
 \frac{k_{w} T_{2t} \Delta } {w_{2}+1/\alpha_{e}+1/\alpha_{c}}\nonumber\\
     & &[L(E_{z}-E_{tn}) \frac{T_{d}}{T_{d}+T_{2t}}+
L(E_{z}+ \hbar w -E_{tn}) \frac{T_{a}}{T_{a}+T_{2t}} + \nonumber\\
     & &L(E_{z}-\hbar w - E_{tn}) \frac{T_{e}}{T_{e}+T_{2t}}] dE_{z}
\end{eqnarray}
\par
\section{RESULTS}
 We have analyzed the effect of an external laser in the far infrared
regime on
the magnetotunneling current density through a semiconductor double
barrier. The magnetic field and the electromagnetic field are applied in
the configuration shown in fig. 1.
In fig 2a the coherent magnetocurrent density
 is represented as a function of the
external bias for a magnetic field of 24 T
 and in the presence of an external electromagnetic field with a
frequency of 10.3 meV and an intensity 5 $ 10^{4} $ V/m .
 In this case, only one LL contributes to the tunneling current and
 the analysis
  of the effect of the light on the current can be done in a simpler
way than in the case where more LL participate in the current.
The current difference between the case where there is a laser present
and where there is no light applied to the heterostructure is
represented in fig 2b.  In this case we observe a main peak which appears
for  smaller bias than the corresponding to the threshold bias for the
magnetotunneling current with no light present. As the bias increases
the current difference decreases and becomes negative. There is also a
small positive and a negative structure for higher bias  and as the bias
corresponding to the cut off of the current is reached
 there is an additional
positive contribution to the current difference .
 These features can be schematically  explained
in figure 3: for small bias the resonant
state in the well with energy corresponding to the first LL is higher in
energy than the Fermi energy in the emitter.
As the bias increases there
are electrons close to the Fermi energy which are able to absorb a
photon and tunnel resonantly from the first Landau level in the emitter
with Landau level index conservation , therefore  the threshold bias
for the current is smaller
than the corresponding one for no light present (it moves twice the
photon energy) and there is a positive peak in the current difference .
For higher bias
the first Landau level in the well crosses the Fermi energy in the
emitter and the current difference becomes negative abruptly due to the
fact that the electrons in the emitter have the  possibility to absorb a
photon and it reduces the number of electrons efficient to tunnel
resonantly. For higher bias there are absorption ,
emission and direct tunneling processes whose
combinations give the positive structure observed.
 As the bias increases
and the energy of the resonant state in the well for the first LL lyes
one photon higher than the conduction band bottom of the emitter
the electrons have a probability
to emit a photon below the bottom of the conduction
band and the resonant current
is reduced. ( it corresponds to the small negative contribution to the
current difference for high bias)
 Once the resonant state crosses the bottom of the
conduction band there are electrons in the emitter which can emit a
photon and tunnel resonantly, therefore there is a positive peak in the
current difference and the current cut off moves to higher bias
. \\
\par
As the magnetic field decreases there are more Landau levels which
contribute coherently to the current. In fig. 4a the current as a
function of the external bias is represented for a field of 8 Tesla. We
observe four LL which contribute to the current. The photon frequency is
 6.9 meV ( one half of the cyclotron frequency). The current difference is
shown in figure 4b. In this case  the main peak in the current difference
due to the effect of the light
appears at different bias for the
different Landau levels and the contribution
at the cut off is added up for the four levels. In fig. 4c. the current
difference has been drawn separately for each Landau level , for the
same case as in fig. 4b . If one now changes the
photon frequency to the same value as the cyclotron one for the same
magnetic field (8T)  the current difference
changes dramatically and the main contribution comes from the peak at
the threshold bias and an additional narrow structure in this region of
bias
(fig. 5a) . For higher bias the additional
 features to the current difference
are much  smaller  in intensity than in the
previous case (fig. 4b) .
 The reason for this difference between both cases is not
only the change of the bias threshold and cut off of the current due to
the difference of photon frequencies (the threshold bias is lower for
higher photon frequencies and the cut off bias is larger for higher
photon frequencies) but also is due to the fact that
when the ratio of the cyclotron frequency
to the photon frequency is one, there are absorption and emission
processes taking place for electrons coming from different Landau levels
which compensate each other. This feature can be observed in fig. 5b ,
 where the
contribution to the current density coming from each LL is represented.
Due to this compensation it is possible to control the the effect of the
light on the magnetocurrent by tuning the ratio between the cyclotron
and the photon frequency.\\
\par
 As we have already discussed, in order to see the total effect of the light
on the magnetocurrent the sequential contribution to the current and the
modifications it suffers due to the light should be analyzed.
We have evaluated the sequential magnetotunneling current for the same
cases as the coherent one: In fig. 6a and 6b the current density  and the
current density  difference are represented
for a magnetic field of 24 Tesla and an electromagnetic field of energy
10.3 meV (one fourth of the cyclotron energy).  The structure
observed in the current difference can be understood in the same footing
as in the coherent case.\\
As the magnetic field is decreased to 8 Tesla
 the four peaks in the  sequential
 magnetocurrent indicates the participation of four
Landau levels (fig.7a). The current difference for a laser field of 6.9
meV (one half of the cyclotron frequency) is represented (fig.7b and
7c) and similar features as in the coherent case are observed but
comparing with the coherent case (fig.4b) we obtain that the effect of
the light is smaller for the sequential current than for the coherent
one in spite of the fact that the current density due to both
contributions are of the same order. When the frequency
of the applied laser is the same as the cyclotron frequency (the photon
energy is 13.8 meV) there are again compensations in the current
difference coming from different Landau levels
( figures 8a and 8b) and the light affects mainly the current density at
the threshold and the cut off bias.\\
\par
 In summary, the observed feature coming from our
calculations is that the coherent contribution to the magnetotunneling
current is more affected by light
 than the sequential one  and that the effect of a laser on both
contributions to the
current can be controlled and modified by tuning the ratio between the
cyclotron and the photon frequency. A drastic modification
is observed
 when this ratio becomes one. \\
\par
There is an additional effect which occurs when the electrons
tunnel sequentially and which can also be externally modified by an
external electromagnetic field.
It is the fact that when sequential tunneling takes place, the electrons
which have tunneled coherently through the first barrier conserving the
Landau level index in the process, relax into the well and tunnel
coherently in a subsequent process through the second barrier.
Therefore the number of the Landau levels at the emitter contributing
to the current can be different than the corresponding to the Landau
levels
in the well participating in the  tunneling current through the second
barrier, and those numbers can be modified applying an external laser.
It can be seen in fig. 9a, where at low field (6T) there are many Landau
levels giving current. The dotted line represents the magnetocurrent
when no light is present into the sample
 and the continuous line corresponds
to the case where a laser with a frequency of 7 meV and intensity of
5 $ 10^{6}$ V/m is applied to the sample. We observed that the current
with no light presents a sawtooth profile coming from the participation
of additional Landau levels as the bias increases.
  When the light is switched on, the
current threshold moves to lower bias and there is a three step like
structure between each jump .
For high electromagnetic field intensities the effect of light
can be seen clearly in the current density curve as in fig. 9a.
 In fig, 9.b) we represent the total
 number of Landau levels
partially occupied in the well as a function of
the bias  for both cases: with (continuous line) and with no
light (dotted line). We observe clearly for instance, that around
 0.04 V the second Landau
level begins to be occupied in the case where the
light is present in the sample. As the bias slightly increases
the second Landau level begins  also to contribute to
the current through the second barrier even with no applied light.
 As the
bias increases, the well is discharged and the second Landau level
becomes empty for the case in which there is no light present
 (dotted line) .
Finally, for high bias, the second
 Landau level becomes discharged
 for smaller bias in
the case where there is no light present (dotted line) than in the
presence of a laser (continuous line).
 From the above discussion we conclude that for a
fixed bias the number of Landau levels participating in the  sequential
magnetocurrent
through the second barrier ( which is determined from the Landau level
occupation in the well) can be modified illuminating
the sample.

\section{CONCLUSIONS}
 In this paper we have extended the formalism proposed for photoassisted
tunneling \cite{13,14} to analyze both coherent and sequential
magnetotunneling through a double barrier structure. We have considered
a magnetic field applied parallel to the current and a laser with the
electric field in the same direction as the current. We have observed
firstable that the modification of the current ( for both, sequential and
coherent processes )
 due to the photon field can be controlled by tuning the ratio
between the cyclotron and the photon frequency, turning on that when
this ratio becomes one, there is a compensation in the current
difference coming from different landau levels and this magnitude
changes drastically and most of the structure observed in
the cases  where the cyclotron and the photon frequency are different
is quenched.\\
\par
Other interesting effect is that the light changes the Landau level
occupation and modifies for a fixed bias, the
number of Landau levels
 which participate in the current through the emitter ($J_{1}$) and
also through the collector barrier ($J_{2}$).
 It means that for a given sample and
an external magnetic field, by changing the photon frequency and
intensity of the external applied laser
 is possible to get information
on the Landau level density of states in the
 well.
\par
In conclusion, the effect of the light on
 the magnetotunneling current
has been analyzed for both coherent and sequential
 tunneling processes,
turning on that it is the former the most affected
 by light. This effect
can be controlled by tuning the ratio
 between the two characteristic
frequencies: the cyclotron and photon ones and it
 can be  drastically
modified when this ratio becomes one.
 Very
interesting effects are also expected for smaller systems where the
electron-electron interaction is important and  where
the occupation of the resonant structure
 is crucial for the electron transport.
 This is the aim of a future work. \\
\par
\newpage
\section{ACKNOWLEDGMENTS}
\par
 This work has been supported in part by the Comision Interministerial
de Ciencia y Tecnologia of Spain under contract MAT 94-0982-c02-02 , by
the Comission of the European Communities under contract SSC-CT 90 0201
and by the Acci\'on Integrada Hispano-Alemana HA93-034.\\

\newpage
\begin{figure}
\caption{Particle represented by a plane wave moving along the $z$ direction
crossing a DBS ( well thickness lw=40 \AA, barrier thickness lb=50 \AA) in the
presence of an electromagnetic field polarized in the
$z$ direction and an magnetic field parallel to z.}
\end{figure}
\begin{figure}
\caption{a) Coherent magnetotunneling current density
 assisted by light as a function of V.
( $F=5.10^{4} V/m$, $\hbar w=10.3 meV$ , B=24 Tesla). b)
 Coherent magnetocurrent density difference as a function of V between
 photoassisted magnetocurrent and magnetocurrent without light
present. }
\end{figure}
\begin{figure}
\caption{Schematic drawn of photoassisted tunneling
 processes for increasing bias.}
\end{figure}
\begin{figure}
\caption{a) Same as fig. 2a for
$F= 5. 10^{4} V/m$, $\hbar w= 6.9 meV$ , B= 8 Tesla .
b) Same as fig. 2b for
 $F= 5. 10^{4} V/m$, $\hbar w= 6.9 meV$ , B= 8 Tesla .
c) Coherent magnetocurrent density difference as a
 function of V for each Landau level
separately.}
\end{figure}
\begin{figure}
\caption{a) Coherent magnetocurrent difference for B= 8 Tesla and $\hbar
w=\hbar w_{cyclotron} $ .
b) Same as in a) for each Landau level separately. }
\end{figure}
\begin{figure}
\caption{a) Sequential magnetotunneling current
 density as a function of
voltage
 ($F= 5. 10^{4} V/m$, $\hbar w= 10.3 meV$ , B= 24 Tesla ).
b) Sequential magnetocurrent difference
  as a function of voltage between
sequential light
assisted tunneling  and sequential
  tunneling without light present.
 Same configuration as in a) }
\end{figure}
\begin{figure}
\caption{a) The same as fig. 6 a) for  B= 8 Tesla,
  $\hbar w= 6.9 meV.$
 b) Same as fig. 6 b) for B= 8 Tesla, $\hbar w= 6.9 meV.$
 c) Same as in b) for each Landau level separately.}
\end{figure}
\begin{figure}
\caption{a) Sequential magnetocurrent
 difference (with and without
light) for B= 8 Tesla and $\hbar w=\hbar w_{cyclotron} . $
 b) Same as  a) for each Landau level separately. }
\end{figure}
\begin{figure}
\caption{a)
 Sequential magnetocurrent density as a function of V. B= 6 Tesla, $\hbar
w= 7 meV $, $F= 5. 10^{6} V/m$. Continuous line: light present;
 Dotted line: no light present.
 b) Total number of Landau levels into the well
contributing to the current as a
function of V with (continuous line) and without (dotted line) light.
 c) Landau levels occupation into the well as a function of V. Continuous
line: light present; Dotted line: no light
 present .}
\end{figure}
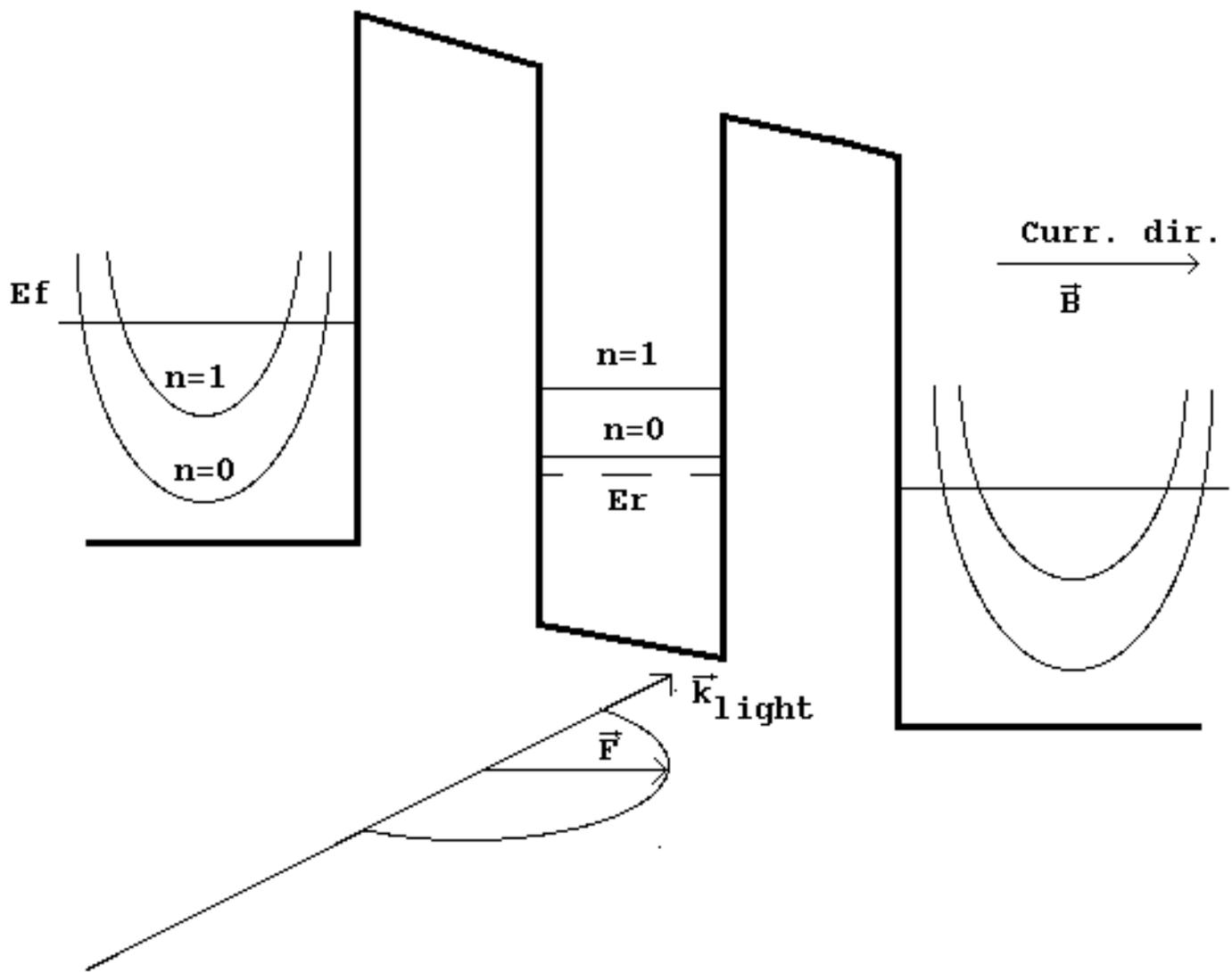
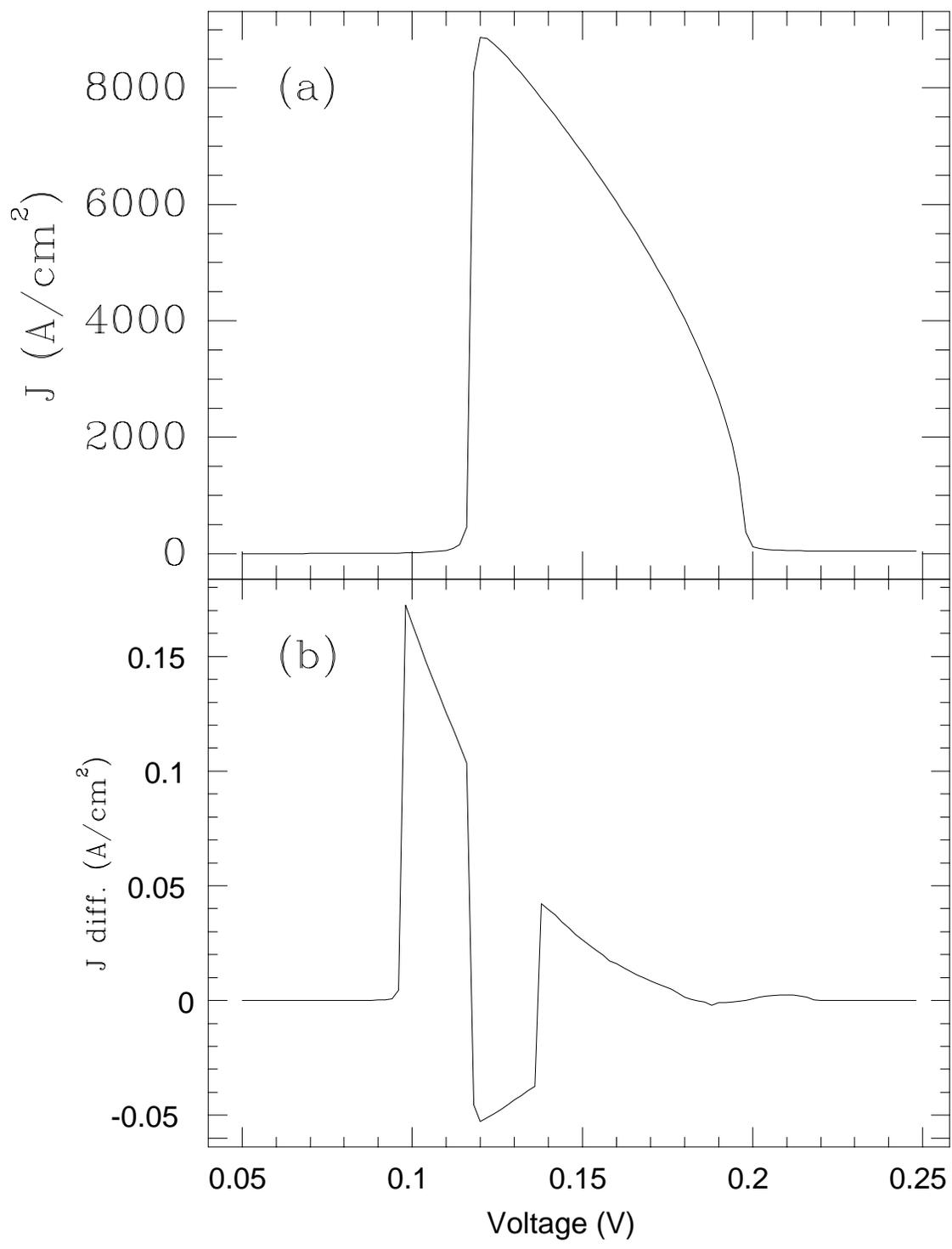
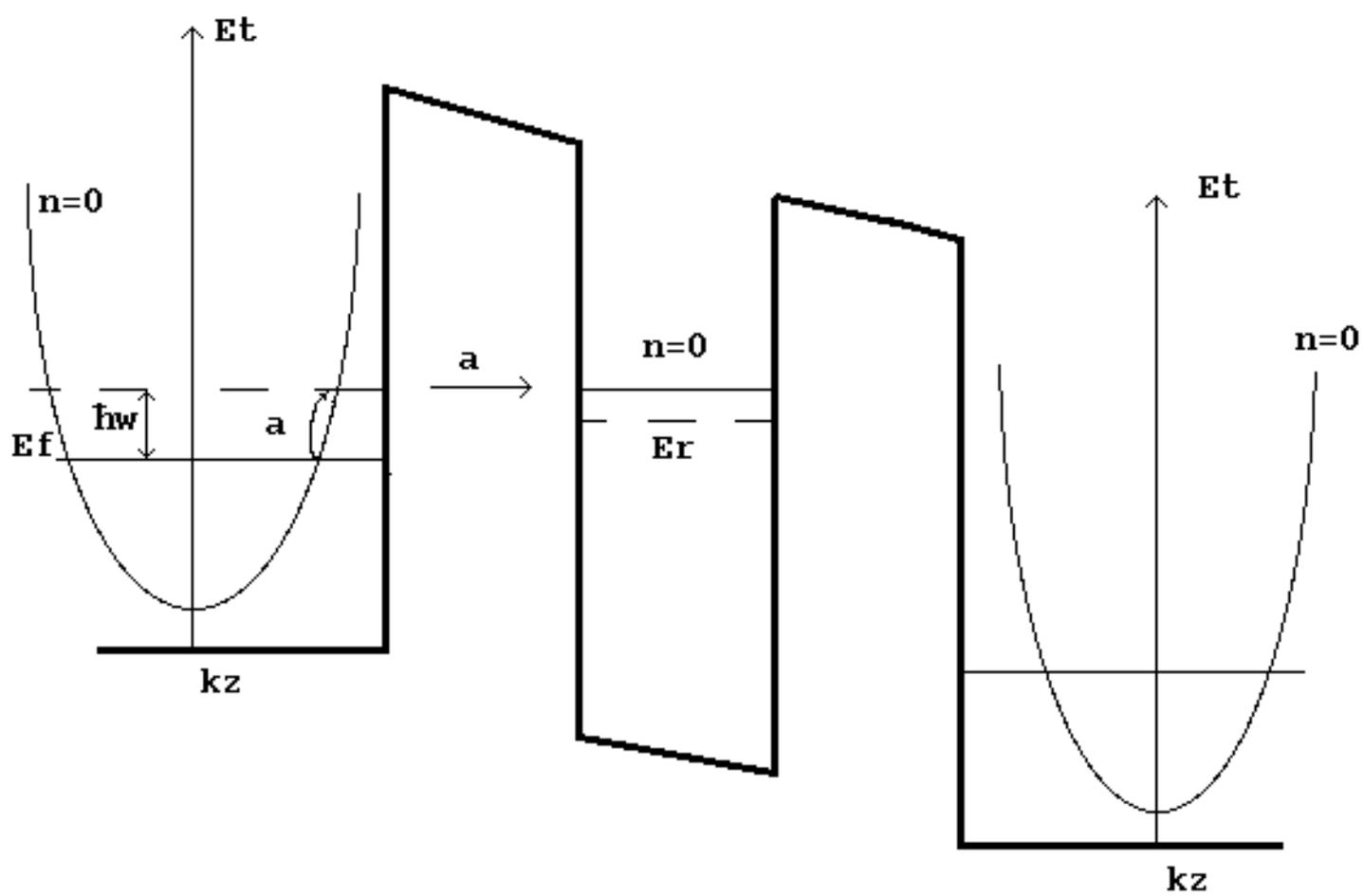
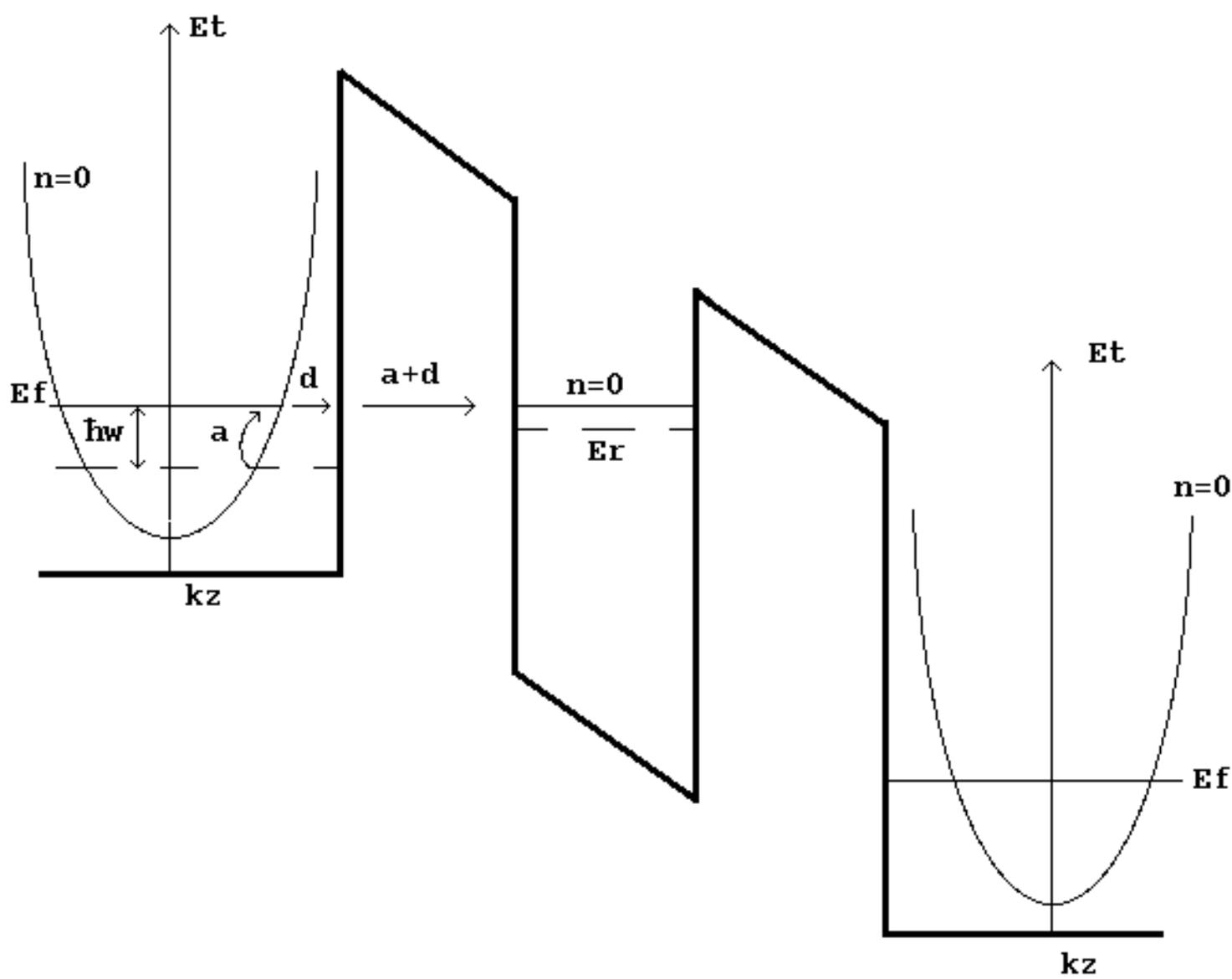
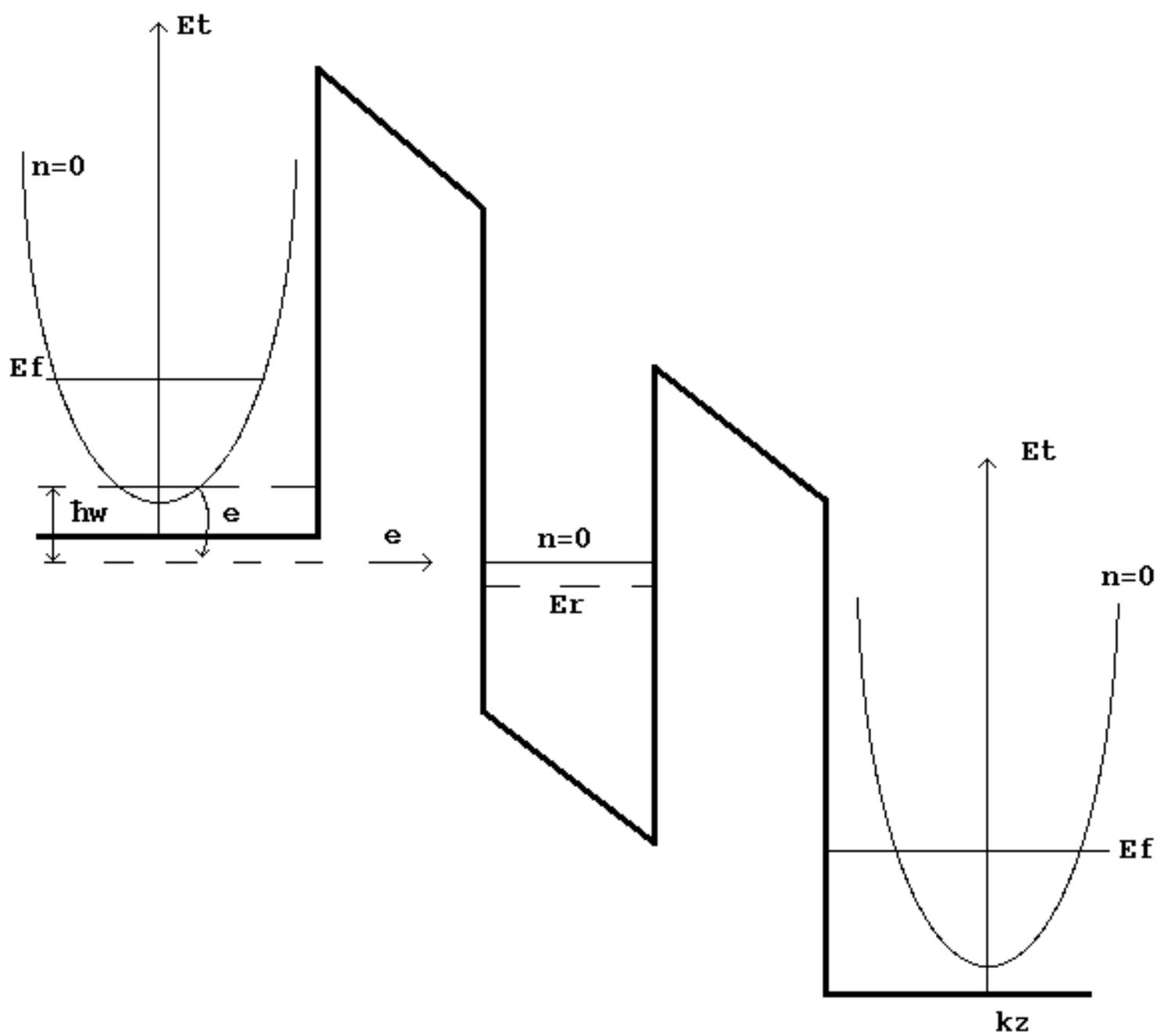
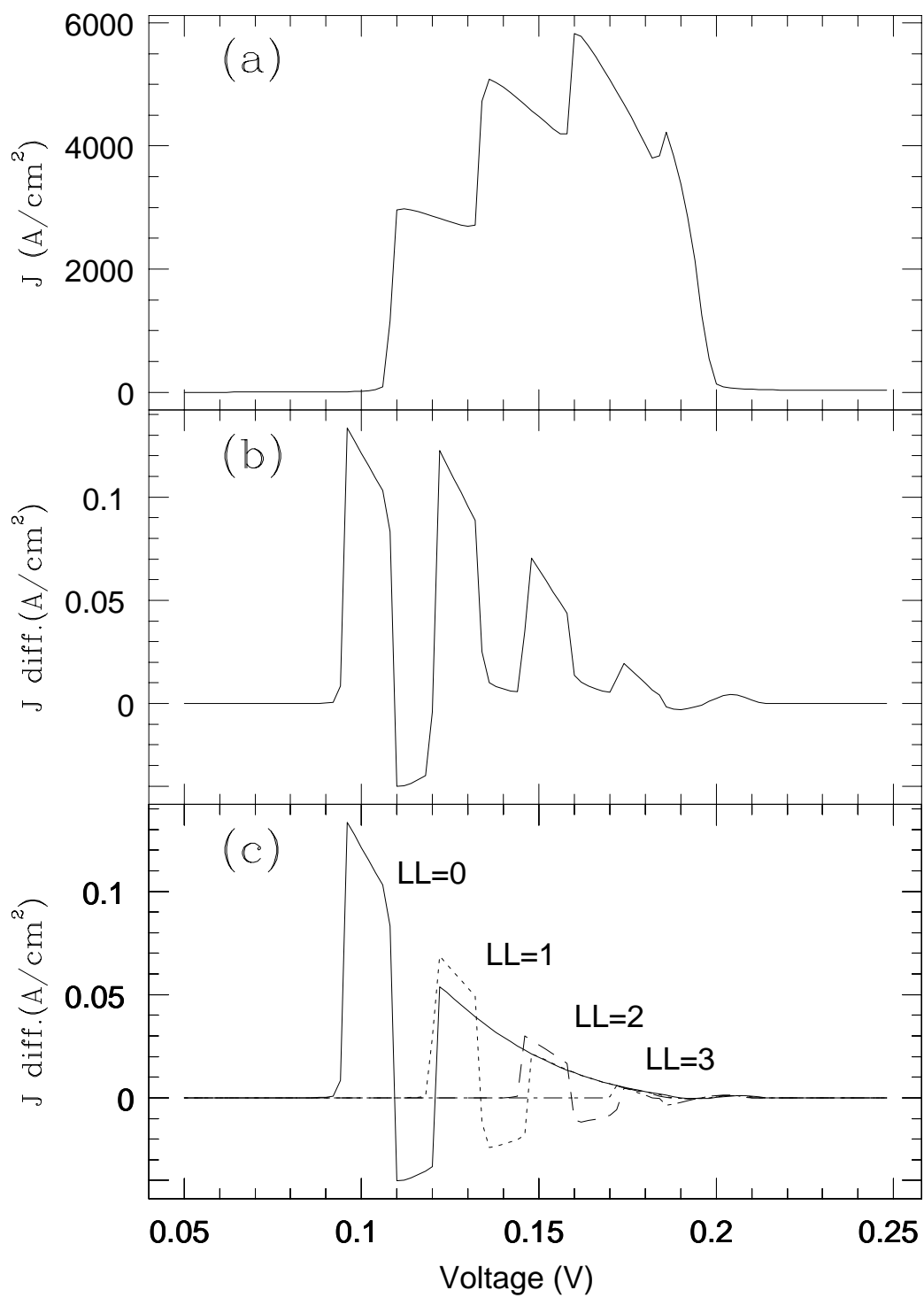
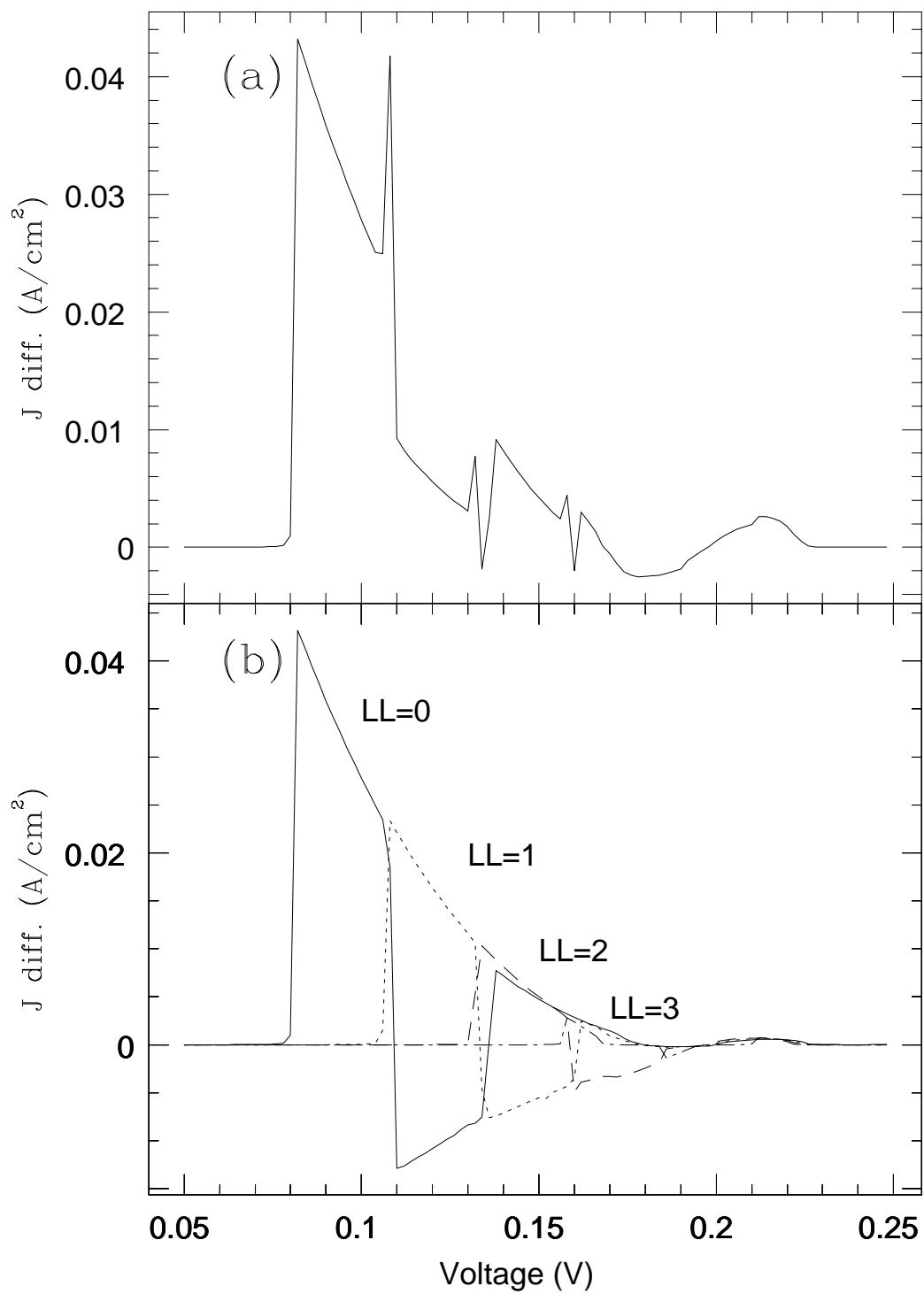
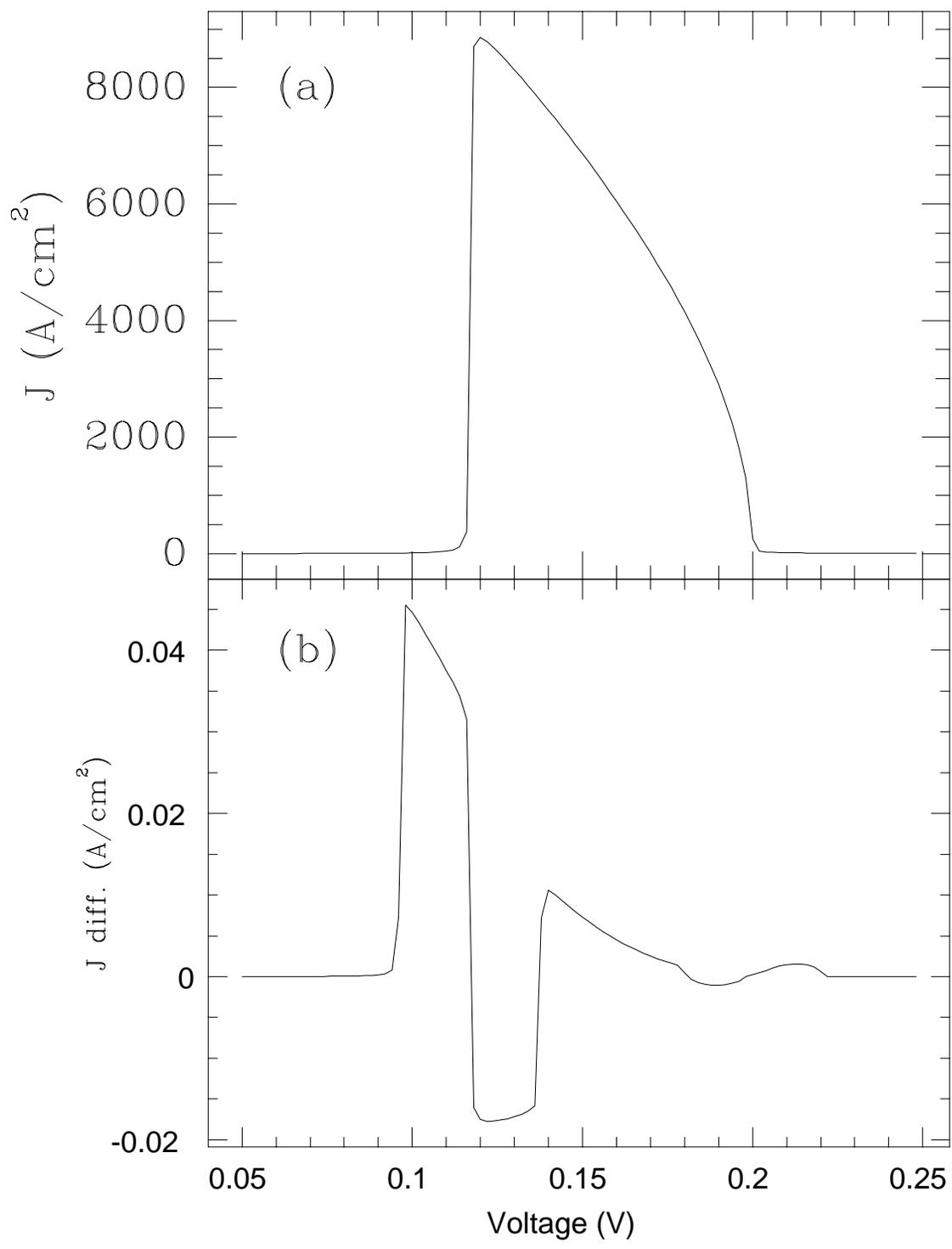
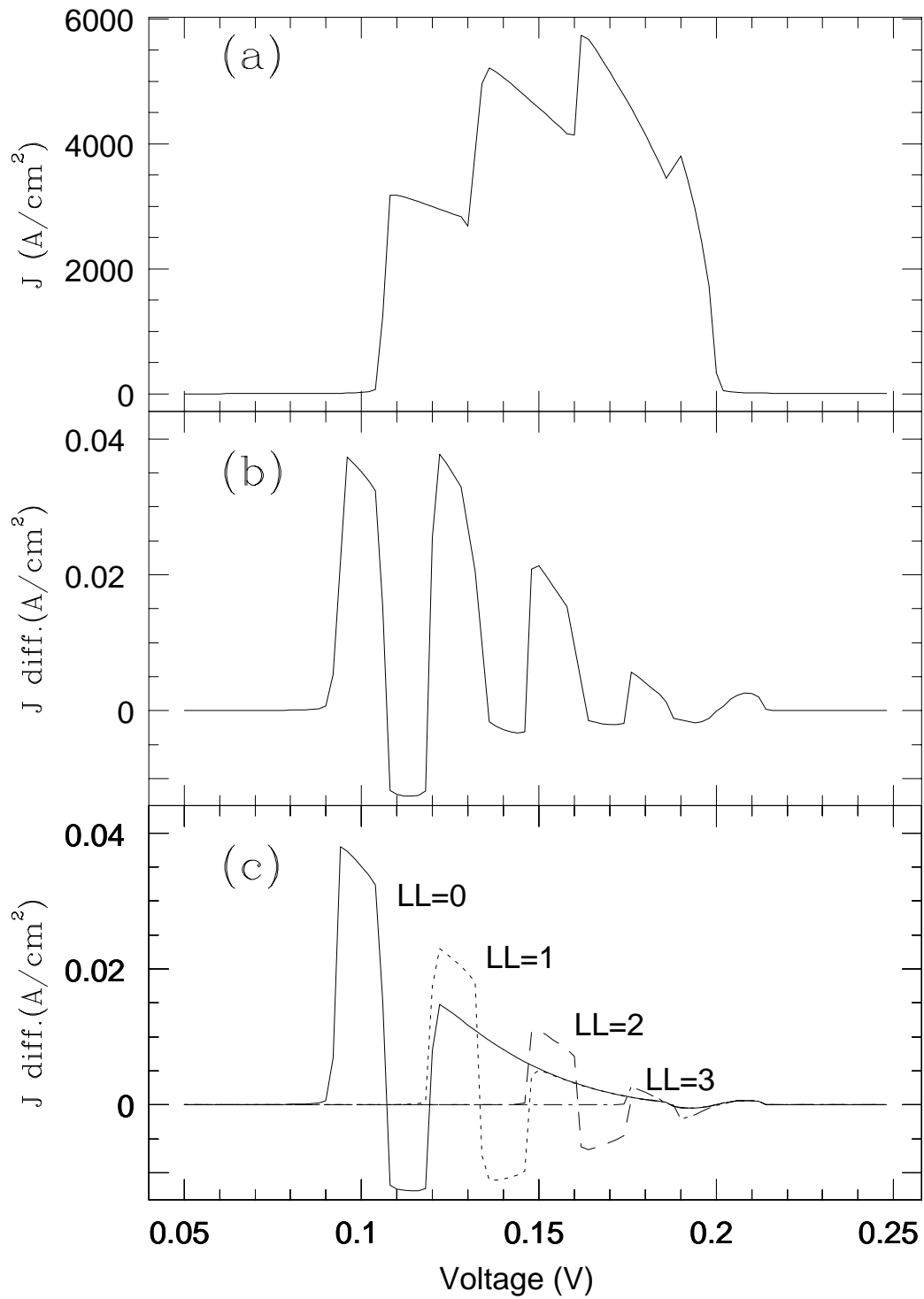
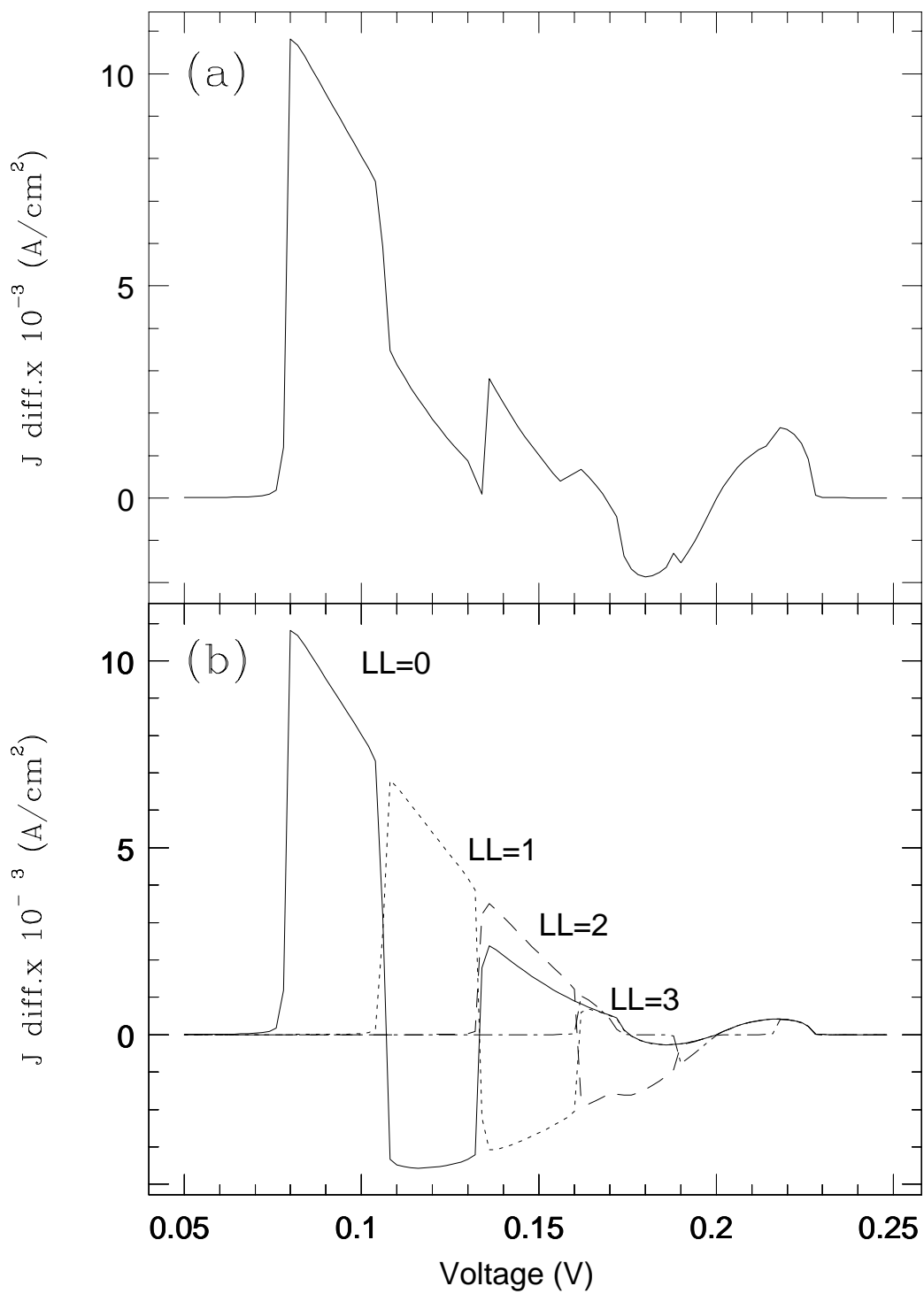
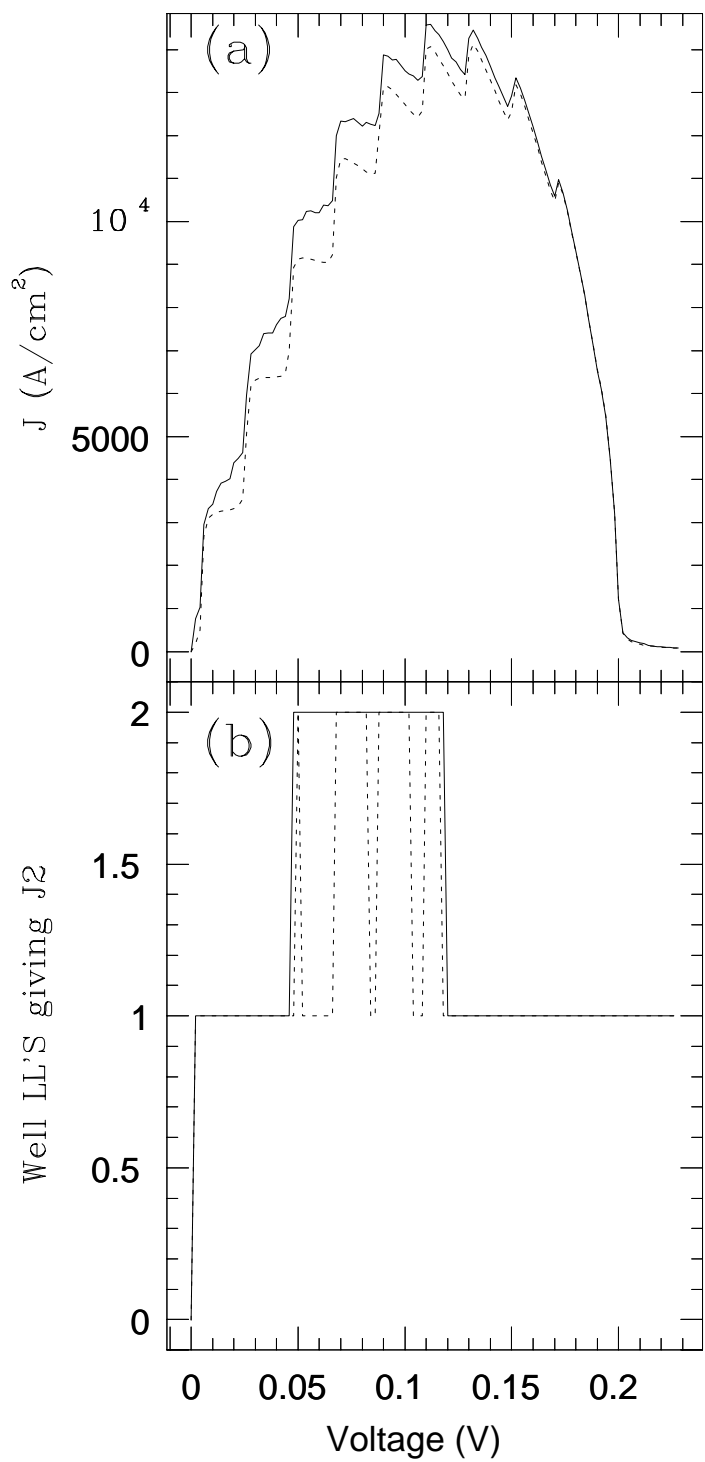
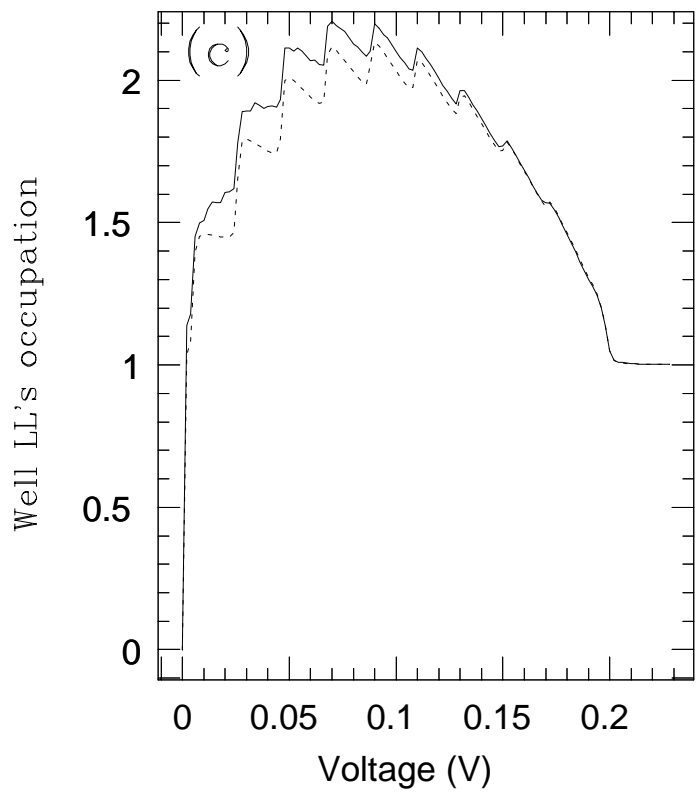
\end{document}